\documentclass[conference, a4paper]{IEEEtran}

\IEEEoverridecommandlockouts                          
\usepackage{units}
\usepackage{graphicx}
\usepackage{multirow}
\usepackage{amsmath,amssymb,latexsym}
\usepackage{rotating}
\usepackage{lettrine}
\usepackage{textcomp}

\usepackage{hyperref}

\usepackage{xcolor}
\usepackage{units}

\usepackage{lipsum}% 
\usepackage[pscoord]{eso-pic}
\newcommand{\placetextbox}[3]{% 
  \setbox0=\hbox{#3}
  \AddToShipoutPictureFG*{
    \put(\LenToUnit{#1\paperwidth},\LenToUnit{#2\paperheight}){\vtop{{\null}\makebox[0pt][c]{#3}}}%
  }%
}%

\title{Mixing Neural Networks and Exponential Moving Averages for Predicting Wireless Links Behavior
\thanks{This work was partially supported by the European Union under the Italian National Recovery and Resilience Plan (NRRP) of NextGenerationEU, partnership on ``Telecommunications of the Future'' (PE00000001 - program ``RESTART''). Computational resources were provided by HPC@POLITO, a project of Academic Computing within the Department of Control and Computer Engineering at the Politecnico di Torino (http://www.hpc.polito.it).}}

\author{
    \IEEEauthorblockN{Gabriele Formis\IEEEauthorrefmark{1}\IEEEauthorrefmark{2}, Stefano Scanzio\IEEEauthorrefmark{1}, Lukasz Wisniewski\IEEEauthorrefmark{3}, and Gianluca Cena\IEEEauthorrefmark{1}}
    \IEEEauthorblockA{\IEEEauthorrefmark{1}National Research Council of Italy (CNR--IEIIT), Italy. \IEEEauthorrefmark{2}Politecnico di Torino, Italy.}
    \IEEEauthorblockA{\IEEEauthorrefmark{3}Institute Industrial IT – inIT
of Technische Hochschule OWL, Germany.}
    Email: gabriele.formis@ieiit.cnr.it, stefano.scanzio@cnr.it, lukasz.wisniewski@th-owl.de, gianluca.cena@cnr.it
    \vspace{-0.2cm}
}

\begin{document}
\placetextbox{0.5}{1}{This is the author's version of an article that has been published.}
\placetextbox{0.5}{0.985}{Changes were made to this version by the publisher prior to publication.}
\placetextbox{0.5}{0.97}{The final version of record is available at \href{https://doi.org/10.1109/ICPS59941.2024.10640038}{https://doi.org/10.1109/ICPS59941.2024.10640038}}%
\placetextbox{0.5}{0.05}{Copyright (c) 2024 IEEE. Personal use is permitted.}
\placetextbox{0.5}{0.035}{For any other purposes, permission must be obtained from the IEEE by emailing pubs-permissions@ieee.org.}%

\maketitle
\thispagestyle{empty}
\pagestyle{empty}

%%%%%%%%%%%%%%%%%%%%%%%%%%%%%%%%%%%%%%%%%%%%%%%%%%%%%%%%%%%%%%%%%%%%%%%%%%%%%%%%
\begin{abstract}
Predicting the behavior of a wireless link in terms of, e.g., the frame delivery ratio, is a critical task for optimizing the performance of wireless industrial communication systems.
This is because industrial applications are typically characterized by stringent dependability and end-to-end latency requirements, which are adversely affected by channel quality degradation.

In this work, we studied two neural network models for \mbox{Wi-Fi} link quality prediction in dense indoor environments. 
Experimental results show that their accuracy outperforms conventional methods based on exponential moving averages, 
due to their ability to capture complex patterns about communications, including the effects of shadowing and multipath propagation, which are particularly pronounced in industrial scenarios.
This highlights the potential of neural networks for predicting spectrum behavior in challenging operating conditions,
and suggests that they can be exploited to improve determinism and dependability of wireless communications, fostering their adoption in the industry.
\end{abstract}

%%%%%%%%%%%%%%%%%%%%%%%%%%%%%%%%%%%%%%%%%%%%%%%%%%%%%%%%%%%%%%%%%%%%%%%%%%%%%%%%

\section{Introduction}
\label{sec:introduction}

The fourth industrial revolution, commonly known as Industry 4.0 \cite{CANAS2021107379,MADDIKUNTA2022100257} is catalyzing a profound transformation in manufacturing processes and communication systems, heralding an era of unprecedented connectivity and automation. 
In this paradigm, wireless networks emerge as linchpins, playing an increasingly pivotal role in facilitating the flexible and scalable interconnection of distributed devices and systems. 
However, this transformative landscape is not without its challenges, and the ever-varying quality of wireless channels \cite{CCNC,ICC,ISCMI,IEEETVT} introduces a layer of complexity shaped by environmental factors such as 
physical obstacles, signal reflections, ambient noise, and electromagnetic interference from other devices operating in the same frequency bands. 
This inherent variability poses a significant threat to the optimal functioning of wirelessly interconnected systems, spanning the realms of industrial operations from real-time communications to high-bandwidth video streaming applications.

At the heart of mitigating above issues lies the critical challenge of satisfactorily predicting the behavior of wireless channels, a cornerstone for ensuring the uninterrupted continuity and reliability of communications within the Industry 4.0 framework. 
The ability to foresee the quality of wireless links and to act proactively when changes are expected holds the key to enhancing the determinism and reliability of communications networks. 
This involves not only optimizing transmission settings, but also identifying and mitigating potential sources of interference through intelligent, adaptive strategies, thereby enabling a more resilient and responsive network management approach.

However, delving into the realm of wireless spectrum prediction reveals a myriad of challenges. 
The non-stationary nature of wireless channels, their inherent complexity, and the paucity of historical data on communication quality pose formidable new research problems. 
Tackling these challenges requires innovative approaches, with the landscape broadly classified into statistical methods and machine learning (ML) techniques \cite{9786784}. 
Statistical methods hinge on historical data about link quality, constructing a stochastic model for predicting the future state of wireless channels. 
On the other hand, ML leverages algorithms to discern the connections among factors influencing wireless channel quality \cite{9945847, 8502636, LEONARDI202257, 2016-TII-WiRed, 9921559, 10034532}, thus enabling more accurate predictions of its future state.
In recent times, ML approaches have emerged as promising contenders, showcasing their prowess in capturing intricate patterns and relationships within data. 
Their ability to navigate the complexity of the wireless spectrum positions them as powerful tools for predicting link quality with a superior degree of accuracy and robustness.

As we witness the ongoing evolution of Industry 4.0, the significance of wireless link quality prediction becomes increasingly apparent, enabling the seamless integration and efficient operation of subsystems interconnected over the air as parts of a heterogeneous industrial network system \cite{9779183}. 
Beyond the realms of academic inquiry, the practical utility of these predictive models is underscored by their potential to revolutionize communication systems, achieving determinism, optimizing reliability, and improving other key performance indicators in the dynamic and interconnected landscape of Industry 4.0.

In this paper, we propose and evaluate a novel class of approaches for link quality prediction in \mbox{Wi-Fi} networks, we named Neural Network for Wireless Channel Quality Prediction (NN-WCQP), which combines a neural network (NN) and a number of linear infinite impulse response (IIR) low-pass filters. 
In practice, the outcomes of transmission attempts are sequentially fed to multiple filters operating in parallel, each of which calculates their exponential moving average (EMA) using a different smoothing factor.
EMA outputs are then provided as input features for the NN, which is in charge to produce the current forecast for the link quality.
The algorithm is assessed using a comprehensive dataset collected from a real-world indoor environment.
The optimal parameters of the NN architecture are determined through a rigorous training phase, 
aimed at minimizing the mean squared error (MSE) on a suitable training dataset. 
Experimental results show that \mbox{NN-WCQP} consistently outperformed traditional methods based on moving averages, showcasing significant improvements in accuracy and robustness to environmental changes. 
Notably, it achieves these enhancements while maintaining a manageable computational complexity, thus allowing for high scalability. 
This research marks a substantial advancement in communication quality prediction in wireless networks, showcasing the potential of NNs to revolutionize their usage in modern industrial systems. 
The following Section~\ref{sec:materiali e metodi} introduces all the concepts related to the NN-WCQP algorithm, including the experimental setup employed to acquire the data for validating the algorithm, 
the EMA model, which is the main building block of the proposed algorithm, and, finally, 
the algorithm itself, together with the performance metrics and the description of the training procedure. 
The algorithm undergoes evaluation in Section~\ref{sec:results}, while Section~\ref{sec:conclusions} presents the concluding remarks.

\section{Materials and Methods}
\label{sec:materiali e metodi}
\subsection{Experimental Setup}

The experimental configuration involved two Linux PCs, each one equipped with two \mbox{Wi-Fi} adapters compliant to IEEE 802.11n (TP-Link TL-WDN4800). 
This resulted in a total of four \mbox{Wi-Fi} stations (STAs), each one associated with a dedicated access point (AP) located approximately three meters apart and tuned on a distinct channel. 
Experiments were performed on non-overlapping channels $1$, $5$, $9$, and $13$ in the $\unit[2.4]{GHz}$ band, strategically chosen to prevent interference.
Each STA/AP pair corresponds to a link operated on one of these channels. Different APs were used because we could not handle multiple \mbox{Wi-Fi} channels on the $\unit[2.4]{GHz}$ band on a single AP.
Frame transmissions occurred nearly simultaneously from the four STAs, triggered by the related 
Linux operating systems, synchronized through the Network Time Protocol (NTP) and enhanced with the \texttt{PREEMPT\_RT} Linux patch to improve soft real-time capabilities.
The payload size of frames was set to $\unit[50]{B}$, which is typical for industrial applications where process data are generally small in terms of size.

The testbed is meant to periodically sample channel conditions with minimal spectrum disruption. 
For this purpose, the device driver of the STAs underwent some modifications: 
transmissions were set to a fixed bit rate ($\unit[54]{Mb/s}$), disabling the Minstrel algorithm; 
automatic frame retransmissions were disabled, so that every frame was sent only once; 
the backoff procedure was turned off, allowing immediate frame transmission when the channel becomes idle; 
the Request to Send/Clear to Send (RTS/CTS) mechanism was disabled, and the same happened to specific IEEE 802.11n features (e.g., frame aggregation), practically downgrading adapter operation to IEEE 802.11g.
Doing so does not pose particular limitations, since the frames sent by the testbed only serve to probe the quality of the considered channels at specific, evenly spaced points in time.
Conversely, nearby interfering nodes (not under our control, neither APs nor STAs) were operating up to Wi-Fi 6.

These modifications were facilitated by the \texttt{ath9k} device driver, which allows an easy modification and configuration of the main \mbox{Wi-Fi} parameters listed above, 
and by the \texttt{SDMAC} framework \cite{8477080,7991945}, which enables relevant information to be transferred from kernel space to user space for database acquisition. 
The application logged (indirectly) the reception status $x_i$ (success/failure) of the $i$-th frame through the event conveyed by the device driver every time the ACK frame related to a data frame was received or the relevant timeout expired.

This testbed aims to characterize the $\unit[2.4]{GHz}$ band in our lab, 
where the wireless spectrum is shared by several \mbox{Wi-Fi} and wireless sensor networks (and the related nodes).
Prior work \cite{ExperimentalEvaluationSeamless2017} shows that, in terms of the amount and pattern of interference, this scenario shares many similarities with what can be seen at the shop floor of industrial plants.
The dynamic nature of the spectrum, influenced by the varying number of active clients (mobiles and notebooks exchanging data with about $50$ APs) as researchers and students move in and out during the day, makes the experimental conditions highly non-stationary.

\subsection{Data Collection}
\label{sub:data_collection}

Data frame transmission in the testbed occurs cyclically with a period of $T_\mathrm{s} = \unit[0.5]{s}$, and the reception of every acknowledgment (ACK) frame is meticulously logged. 
Outcomes are categorized as success ($x_i = 1$) when the ACK frame associated with the $i$-th data frame is correctly received by the sender node. 
Conversely, outcomes are labeled as failure ($x_i = 0$) in cases where either the data frame or the ACK frame are corrupted and the transmission timeout expires. 
Acquisition intentionally mirrors the perspective of the application executing in the sender node, which only indirectly becomes aware of transmission errors.
In essence, for any of the considered channels this experimental setup generates an ordered sequence of outcomes $D = (x_1, ..., x_i, ..., x_{|D|})$, where every outcome $x_i$ is a binary value.

\begin{figure*}[t]
    \begin{center}
    \includegraphics[width=2.00\columnwidth]{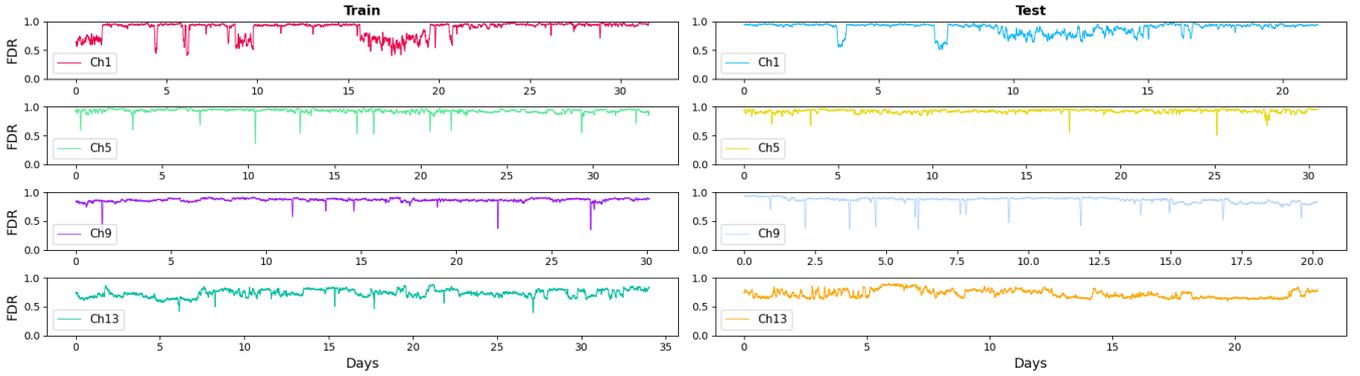}
    \end{center}
    \vspace{-0.2cm}
    \caption{Timing diagrams of the measured FDR (target) for the training and test datasets related to the four considered channels (1, 5, 9, and 13)}
    \vspace{-0.2cm}
	\label{fig:FDR}
\end{figure*}

From the comprehensive logs acquired through the previously described experimental procedure, training and test databases were extracted without any specific criteria, in order to obtain results of general validity. 
Fig.~\ref{fig:FDR} presents eight timing diagrams depicting the frame delivery ratio (FDR) for these databases. 
Training databases are displayed on the left column, while the right column pertains to test databases. 
Rows correspond to channels $1$, $5$, $9$, and $13$. 
These diagrams illustrate (indirectly, by means of the FDR) the variations in the interference levels experienced on channels and the related interference patterns.

All channels considered, experimental data used for training and testing covered $121$ and $99$ days, respectively. 
Overall, databases encompassed $220$ days, equivalent to approximately seven months.
Although our work focused exclusively on \mbox{Wi-Fi}, the proposed approach can be readily extended to any wireless communication technologies that support confirmed transmission services.

\subsection{Exponential Moving Average (EMA) Model}

The EMA model is a smoothing technique designed to assign exponentially decreasing weights to past observations. It has proven to be effective in real problems in comparison with other techniques \cite{WFCS2023}.
This implies that recent data points carry more influence than older ones, allowing the EMA to effectively capture the most recent trends in the data. 
Its applicability shines in scenarios involving time series characterized by non-stationarity, as it adapts dynamically to shifts in the underlying patterns.

In the context of this analysis, EMA is employed to estimate the current FDR by considering the recent history of successful and failed transmissions.
The FDR at time step $i$ is expressed as
\begin{eqnarray}
\label{eq:EMA}
    y^{\mathrm{EMA}}_i = \alpha x_i + (1 - \alpha) y^{\mathrm{EMA}}_{i-1}, 
\end{eqnarray}
where 
$\alpha$ is the smoothing factor ($0 < \alpha \leq 1$), a coefficient 
that fully characterizes the EMA model (i.e., filter),
and $x_i$ is the transmission outcome at time step $i$.

Coefficient $\alpha$ plays a crucial role in determining how responsive the EMA is to changes in channel quality. 
Higher $\alpha$ values weight recent observations more, making the EMA more responsive to changes in the FDR
but also
more susceptible to noise. 
Conversely, low $\alpha$ values provide more weight to older observations, yielding higher stability and lower susceptibility to noise. 
In the context of this work, the optimal value for $\alpha$ (termed $\alpha^*$) was estimated as the one that minimizes the FDR prediction error on a training database. 
More details on its evaluation can be found in \cite{INDIN2023}.
The EMA stands out as a straightforward yet effective technique for smoothing time series data, offering a versatile tool for estimating the current FDR in various applications.
For further discussion on the EMA and its applications in time series analysis applied to wireless communication, see \cite{WFCS2023, INDIN2023}.

\begin{figure*}[t]
    \begin{center}
    \includegraphics[width=2\columnwidth]{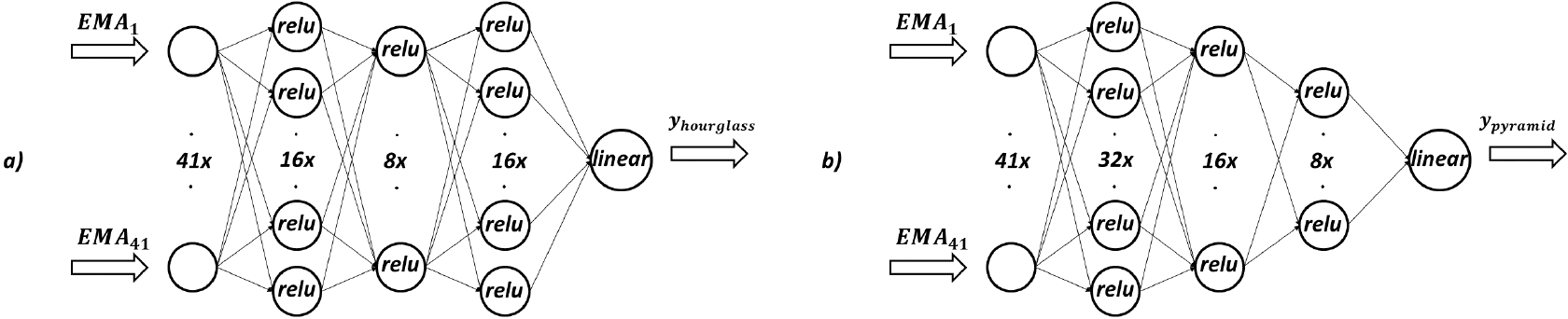}
    \end{center}
    %\vspace{-0.2cm}
    \caption{The two NN architectures we analyzed in this work (four layers each, with the specified number of neurons): hourglass (a) and pyramid (b)}
    %\vspace{-0.2cm}
	\label{fig:NN}
\end{figure*}

\subsection{NN of EMA Models}

In recent years, NNs \cite{NNPred} have been increasingly and profitably used to address self-learning problems, 
such as time series prediction, which is the primary topic covered in this work.
A fundamental capability of NNs lies in their aptitude for learning from data, a process known as training. During training, adjustments are made to the weights of the connections between neurons and biases. 
The overall objective is to minimize the error between the NN's predicted output and the actual target value. Through iterative refinement of weights and biases, NNs progressively enhance their capacity to make accurate predictions.

In the domain of wireless communication, NNs of type multilayer perceptron (MLP) have proven suitable for predicting channel quality and optimizing network performance \cite{10295470} \cite{Short-Term-Prediction}. 
Notably, in many research works, the \textit{hourglass} \cite{Hourglass} and \textit{pyramid} \cite{Pyramid} models have garnered attention for their efficacy and adaptability. 
Besides MLP, we also considered other NN architectures, such as long short-term memory (LSTM), 
but in the experimental campaigns we have carried out so far they have shown worse results than MLP, 
and for this reason they have not been included in this work. 

At any time step $i$, the input of the NN in NN-WCQP is provided by an array 
$\overline{I}_i = (y^{\mathrm{EMA}_1}_i, y^{\mathrm{EMA}_2}_i,\cdots, y^{\mathrm{EMA}_n}_i)$ 
that represents the outputs of $n$ EMA models operating in parallel and characterized by different values of $\alpha$.
Such coefficients are obtained by multiplying and dividing the optimal value $\alpha^*$ by multiples of $\sqrt{2}$. 
In practice, a number $n=41$ of inputs were computed by separately applying \eqref{eq:EMA} to the following sequence of $\alpha$ values:
\begin{equation}
    \Bigg(\frac{\alpha^*}{20\sqrt{2}}, \frac{\alpha^*}{19\sqrt{2}},...,\alpha^*,...,19\sqrt{2}\alpha^*,20\sqrt{2}\alpha^*\Bigg).
\end{equation}

The NN implements a non-linear function that, given the input vector $\overline{I}_i$ and the vector of parameters $\overline{w}$ that characterizes the NN, returns the FDR prediction made by NN-WCQP
\begin{equation}
    y^{\operatorname{NN-WCQP}}_i = f(\overline{w}, \overline{I}_i).
\end{equation}
The value we selected for $n$ is large enough to cope with both slow and fast variations of the channel quality.
The training procedure of the NN model permits to combine EMA models at best, to maximize prediction accuracy.

As depicted in Fig.~\ref{fig:NN}, two NN architectures were tested.
The \textit{hourglass} (Fig.~\ref{fig:NN}.a) is characterized by an initial decrease in the number of neurons per layer, followed by a subsequent increase. 
The bottleneck in the NN structure permits to reduce the dimensionality of the inputs, 
and implements a sort of principal component analysis (PCA) supervised by the NN training algorithm.
In contrast, the \textit{pyramid} (Fig.~\ref{fig:NN}.b) adopts a hierarchical architecture that resembles a pyramid. Information is processed at multiple levels, with each layer focusing on specific aspects of the input data through the progressive reduction of the number of neurons.
The architectures of the two NN models analyzed in this work, including the number of layers, the number of neurons, and their type, are reported in 
the figure.
Such configurations are those that provided the best results among the many experiments we performed on our databases.
Both the hourglass and pyramid models have demonstrated promising results in predicting wireless link quality, surpassing traditional methods in various scenarios.

\begin{table*}[t]
  \caption{Statistics about the prediction error of the EMA model for the four considered channels}
    \vspace{-0.3cm}
  \label{tab:res_EMA}
  \footnotesize
  \begin{center}
    \tabcolsep=0.145cm
    \def\arraystretch{1.10}
    \begin{tabular}{cccc|ccc|cccccc|cccc}
    Test & Prediction & Training & Model  & $\mu_{e^2}$ & $e^2_{\mathrm{p}_{95}}$ & $e^2_{\mathrm{max}}$ & $\mu_{|e|}$ & $\sigma_{|e|}$ & ${|e|}_{\mathrm{p}_{90}}$ & ${|e|}_{\mathrm{p}_{95}}$ & ${|e|}_{\mathrm{p}_{99}}$ & ${|e|}_{\mathrm{max}}$ & ${e}_{\mathrm{min}}$ & ${e}_{\mathrm{p}_{5}}$ & ${e}_{\mathrm{p}_{95}}$ & ${e}_{\mathrm{max}}$ \\
    channel & model & channel & parameters & \multicolumn{3}{c|}{$[\cdot 10^{-3}]$} & \multicolumn{6}{c|}{[\%]} & \multicolumn{4}{c}{[\%]} \\
    \hline
    \multirow{3}{*}{$\mathrm{ch1}$} 
        & EMA & $\mathrm{ch1}$ & $\alpha^*=0.000900$ & 2.03 & 10.58 & 111.53 & 2.64 & 3.65 & 6.92 & 10.29 & 17.57 & 33.40 & -31.35 & -6.73 & 7.16 & 33.40\\
        & EMA & $\mathrm{all}$ & $\alpha^*=0.000325$ & 2.10 & 10.74 & 114.37 & 2.68 & 3.71 & 6.97 & 10.36 & 18.14 & 33.82 & -32.11 & -6.73 & 7.25 & 33.82\\          
        & EMA & $\overline{\mathrm{ch1}}$ & $\alpha^*=0.000150$ & 2.56 & 13.54 & 115.81 & 2.96 & 4.11 & 8.02 & 11.64 & 20.70 & 34.03 & -33.64 & -8.28 & 7.71 & 34.03\\
    \hline
    \multirow{3}{*}{$\mathrm{ch5}$} 
        & EMA & $\mathrm{ch5}$ &  $\alpha^*=0.000085$ & 1.15 & 2.43 & 190.43 & 1.89 & 2.81 & 3.90 & 4.93 & 12.22 & 43.64 & -43.64 & -3.61 & 4.16 & 15.64 \\
        & EMA & $\mathrm{all}$ & $\alpha^*=0.000325$ & 1.24 & 2.63 & 200.70 & 1.81 & 3.03 & 3.62 & 5.14 & 14.65 & 44.80 & -44.80 & -3.49 & 3.74 & 36.84\\
        & EMA & $\overline{\mathrm{ch5}}$ & $\alpha^*=0.000475$ & 1.31 & 2.62 & 212.68 & 1.80 & 3.14 & 3.56 & 5.11 & 15.26 & 46.12 & -44.95 & -3.50 & 3.61 & 46.12\\
    \hline
    \multirow{3}{*}{$\mathrm{ch9}$} 
        & EMA & $\mathrm{ch9}$ &  $\alpha^*=0.000090$ & 3.68 & 10.03 & 248.20 & 2.65 & 5.46 & 5.54 & 10.01 & 31.44 & 49.82 & -49.82 & -4.56 & 6.01 & 21.43  \\
        & EMA & $\mathrm{all}$ & $\alpha^*=0.000325$ & 4.43 & 18.88 & 251.93 & 2.6o & 6.12 & 4.44 & 13.74 & 34.45 & 50.19 & -50.19 & -3.62 & 5.03 & 45.90 \\    
        & EMA & $\overline{\mathrm{ch9}}$ & $\alpha^*=0.000450$ & 4.74 & 18.64 & 297.14 & 2.61 & 6.37 & 3.95 & 13.65 & 37.61 & 54.51 & -50.43 & -3.47 & 4.38 & 54.51\\
    \hline
    \multirow{3}{*}{$\mathrm{ch13}$} 
        & EMA & $\mathrm{ch13}$ & $\alpha^*=0.000500$ & 1.22 & 5.82 & 41.55 & 2.42 & 2.53 & 5.51 & 7.63 & 12.29 & 20.38 & -18.45 & -5.31 & 5.76 & 20.38 \\
        & EMA & $\mathrm{all}$ & $\alpha^*=0.000325$ & 1.22 & 5.82 & 40.55 & 2.42 & 2.52 & 5.55 & 7.63 & 12.16 & 20.14 & -17.44 & -5.27 & 5.85 & 20.14 \\
        & EMA & $\overline{\mathrm{ch13}}$ & $\alpha^*=0.000275$ & 1.23 & 5.76 & 40.72 & 2.43 & 2.52 & 5.60 & 7.59 & 12.11 & 20.18 & -16.90 & -5.31 & 5.92 & 20.18\\
        \hline
    \end{tabular}
    \end{center}
\end{table*}

\begin{table*}[t]
  \caption{Statistics about the prediction error of NN-WCQP models (hourglass and pyramid) for the four considered channels}
  \label{tab:res_NN}
  \footnotesize
  \begin{center}
    \tabcolsep=0.145cm
    \def\arraystretch{1.10}
    \begin{tabular}{cccc|ccc|cccccc|cccc}
    Test & Prediction & Training & Model  & $\mu_{e^2}$ & $e^2_{\mathrm{p}_{95}}$ & $e^2_{\mathrm{max}}$ & $\mu_{|e|}$ & $\sigma_{|e|}$ & ${|e|}_{\mathrm{p}_{90}}$ & ${|e|}_{\mathrm{p}_{95}}$ & ${|e|}_{\mathrm{p}_{99}}$ & ${|e|}_{\mathrm{max}}$ & ${e}_{\mathrm{min}}$ & ${e}_{\mathrm{p}_{5}}$ & ${e}_{\mathrm{p}_{95}}$ & ${e}_{\mathrm{max}}$ \\
    channel & model & channel & parameters & \multicolumn{3}{c|}{$[\cdot 10^{-3}]$} & \multicolumn{6}{c|}{[\%]} & \multicolumn{4}{c}{[\%]} \\
    \hline
    \multirow{6}{*}{$\mathrm{ch1}$} 
        & Hourglass & $\mathrm{ch1}$ & $\alpha^*=0.000900$ & 2.01 & 10.59 & 120.31 & 2.78 & 3.52 & 7.15 & 10.29 & 16.51 & 34.69 & -34.69 & -6.70 & 7.50 & 33.54\\ 
        & Pyramid & $\mathrm{ch1}$ & $\alpha^*=0.000900$ & 2.07 & 11.05 & 120.16 & 2.80 & 3.59 & 7.30 & 10.51 & 16.97 & 34.66 & -34.66 & -6.67 & 7.85 & 33.94\\     
        & Hourglass & $\mathrm{all}$ & $\alpha^*=0.000325$ & 1.90 & 9.89 & 109.07 & 2.67 & 3.44 & 6.88 & 9.94 & 15.95 & 33.03 & -32.03 & -7.35 & 6.44 & 33.03\\
        & Pyramid & $\mathrm{all}$ & $\alpha^*=0.000325$ & 1.87 & 9.64 & 106.21 & 2.65 & 3.42 & 6.80 & 9.82 & 15.93 & 32.59 & -32.00 & -7.21 & 6.39 & 32.59\\
        & Hourglass & $\overline{\mathrm{ch1}}$ & $\alpha^*=0.000150$ & 2.45 & 12.61 & 111.35 & 2.86 & 4.04 & 7.19 & 11.23 & 21.34 & 33.37 & -31.23 & -8.91 & 5.96 & 33.37\\
        & Pyramid & $\overline{\mathrm{ch1}}$ & $\alpha^*=0.000150$ & 3.27 & 14.25 & 111.97 & 3.50 & 4.52 & 9.83 & 11.94 & 22.06 & 33.46 & -32.51 & -9.82 & 9.82 & 33.46\\
    \hline
    \multirow{6}{*}{$\mathrm{ch5}$} 
        & Hourglass & $\mathrm{ch5}$ &  $\alpha^*=0.000085$ & 0.90 & 1.71 & 195.43 & 1.68 & 2.50 & 3.19 & 4.14 & 10.42 & 44.21 & -44.21 & -3.68 & 2.81 & 22.14 \\
        & Pyramid & $\mathrm{ch5}$ &  $\alpha^*=0.000085$ & 0.93 & 1.79 & 191.64 & 1.72 & 2.52 & 3.23 & 4.23 & 10.45 & 43.78 & -43.78 & -3.82 & 2.83 & 30.07 \\
        & Hourglass & $\mathrm{all}$ & $\alpha^*=0.000325$ & 0.99 & 2.02 & 194.05 & 1.70 & 2.64 & 3.34 & 4.50 & 11.20 & 44.05 & -44.05 & -3.11 & 3.51 & 35.95\\
        & Pyramid & $\mathrm{all}$ & $\alpha^*=0.000325$ & 0.97 & 2.03 & 194.48 & 1.69 & 2.61 & 3.35 & 4.51 & 10.95 & 44.10 & -44.10 & -3.12 & 3.52 & 35.94\\
        & Hourglass & $\overline{\mathrm{ch5}}$ & $\alpha^*=0.000475$ & 1.07 & 2.09 & 196.75 & 1.72 & 2.78 & 3.36 & 4.57 & 12.40 & 44.36 & -44.36 & -3.13 & 3.54 & 37.72\\
        & Pyramid & $\overline{\mathrm{ch5}}$ & $\alpha^*=0.000475$ & 1.06 & 2.15 & 195.53 & 1.71 & 2.77 & 3.39 & 4.63 & 12.32 & 44.22 & -44.22 & -3.13 & 3.58 & 38.52\\
    \hline
    \multirow{6}{*}{$\mathrm{ch9}$} 
        & Hourglass & $\mathrm{ch9}$ &  $\alpha^*=0.000090$ & 2.90 & 3.18 & 259.60 & 2.32 & 4.87 & 3.76 & 5.64 & 29.26 & 50.95 & -50.95 & -4.29 & 3.40 & 43.32  \\
        & Pyramid & $\mathrm{ch9}$ &  $\alpha^*=0.000090$ & 3.12 & 4.88 & 259.47 & 2.66 & 4.92 & 4.69 & 6.99 & 29.77 & 50.94 & -50.94 & -6.12 & 4.06 & 26.59  \\
        & Hourglass & $\mathrm{all}$ & $\alpha^*=0.000325$ & 3.07 & 4.13 & 261.75 & 2.22 & 5.08 & 3.55 & 6.43 & 28.98 & 51.16 & -51.16 & -3.49 & 3.58 & 30.80 \\
        & Pyramid & $\mathrm{all}$ & $\alpha^*=0.000325$ & 3.02 & 3.90 & 260.48 & 2.19 & 5.04 & 3.50 & 6.25 & 28.78 & 51.04 & -51.04 & -3.48 & 3.50 & 31.84 \\
        & Hourglass & $\overline{\mathrm{ch9}}$ & $\alpha^*=0.000450$ & 3.20 & 4.64 & 261.66 & 2.30 & 5.17 & 3.66 & 6.81 & 29.03 & 51.15 & -51.15 & -3.50 & 3.74 & 34.36\\
        & Pyramid & $\overline{\mathrm{ch9}}$ & $\alpha^*=0.000450$ & 3.11 & 4.21 & 278.62 & 2.26 & 5.10 & 3.59 & 6.49 & 28.91 & 52.78 & -52.78 & -3.64 & 3.56 & 30.60\\   
    \hline
    \multirow{6}{*}{$\mathrm{ch13}$} 
        & Hourglass & $\mathrm{ch13}$ & $\alpha^*=0.000500$ & 1.14 & 5.09 & 38.22 & 2.46 & 2.30 & 5.21 & 7.14 & 11.22 & 19.55 & -17.87 & -5.14 & 5.30 & 19.55 \\
        & Pyramid & $\mathrm{ch13}$ & $\alpha^*=0.000500$ & 1.15 & 5.16 & 39.78 & 2.48 & 2.31 & 5.26 & 7.18 & 11.23 & 19.95 & -17.87 & -5.17 & 5.36 & 19.95 \\
        & Hourglass & $\mathrm{all}$ & $\alpha^*=0.000325$ & 1.14 & 4.94 & 42.54 & 2.47 & 2.30 & 5.30 & 7.03 & 11.07 & 20.63 & -17.63 & -5.37 & 5.22 & 20.63 \\
        & Pyramid & $\mathrm{all}$ & $\alpha^*=0.000325$ & 1.14 & 4.96 & 41.03 & 2.47 & 2.31 & 5.30 & 7.04 & 11.10 & 20.25 & -17.48 & -5.37 & 5.21 & 20.25 \\
        & Hourglass & $\overline{\mathrm{ch13}}$ & $\alpha^*=0.000275$ & 1.23 & 4.95 & 41.56 & 2.65 & 2.29 & 5.53 & 7.03 & 10.86 & 20.39 & -17.85 & -6.00 & 4.63 & 20.39\\
        & Pyramid & $\overline{\mathrm{ch13}}$ & $\alpha^*=0.000275$ & 1.23 & 4.97 & 49.55 & 2.65 & 2.29 & 5.52 & 7.05 & 10.94 & 22.26 & -17.86 & -6.00 & 4.61 & 22.26\\
    \hline
    \end{tabular}
    \end{center}
\end{table*}

NN models were trained using the classical backpropagation algorithm on the training database, whose goal is to minimize the MSE between the prediction made by the model and the target, which coincides with the FDR computed on the next $3600$ samples (corresponding to a $30$-minute-wide future interval). 
This is the same objective function as the one employed to compute $\alpha^*$ for EMA.
The optimization process iteratively adjusts the model's weights and biases to enhance prediction accuracy. 
The training process, which was carried out with the \textit{Keras} module of \textit{TensorFlow}, includes $15$ epochs, has a batch size equal to $64$, and uses the \textit{Adam} optimizer. 
The initial learning rate was set to $0.01$ and halved at each epoch.

\subsection{Performance Metrics}

Various metrics were employed in the test phase to assess the accuracy of prediction models. 
One of the most relevant indicators, we will use for comparison in the next section, is the MSE ($\mu_{e^2}$), 
which measures the average squared difference between the model's predictions and the actual FDR values. 
A lower value of $\mu_{e^2}$ indicates higher prediction accuracy, highlighting the model's ability to produce reliable forecasts.

In addition, 
the distribution of errors can also be examined, 
and here come into play the high percentiles (${|e|}_{\mathrm{p}_{95}}$) and maximum (${|e|}_{\max}$) of the \textit{absolute} error. 
For instance, ${|e|}_{\mathrm{p}_{95}}$ provides a critical indication, as it represents the error below which $95\%$ of the predictions lie.
Parallel to them, the mean absolute error (MAE) specifies the average magnitude $\mu_{|e|}$ of prediction errors,
while $\sigma_{|e|}$ is their standard deviation.
Together, they help understanding the overall extent of discrepancies between predictions and reality.
The statistical indexes about the \textit{signed} error, e.g., minimum ($e_{\min}$) and maximum ($e_{\max}$),
outline the range where the majority of errors concentrate,
hence offering an overview of the central tendency of the errors made by the model.

\section{Results}
\label{sec:results}

In this section, the results of our experiments are presented.
We evaluated the accuracy of three different prediction models (EMA, hourglass NN, and pyramid NN)
on the datasets described in Subsection~\ref{sub:data_collection}. 
Every model was trained with three different datasets: 
single channel (e.g., ``$\mathrm{ch1}$'' means the training dataset related to channel $1$), 
all channels (``all'' means that the training datasets of all channels were merged together), and
all channels except one (e.g., ``$\overline{\mathrm{ch1}}$'' means that the training datasets of all channels other than $1$ were merged). 
In the latter case, we excluded the dataset related to the channel 
under test
to make training truly channel-independent.

Table~\ref{tab:res_EMA} reports results for the optimal EMA model (where $\alpha$ is set to $\alpha^*$), 
which provides the best results EMAs can achieve by construction.
They highlight that, after training, this model is generally able to predict the link quality with good accuracy. 
The MSE and MAE are relatively low, and the same holds for the 95- and 99-percentiles. 
However, the maximum error is not negligible.
For example, for channel $1$ the MSE is $2.03 \cdot 10^{-3}$, but the worst-case error is $33.40\%$. 
This suggests that the EMA model was sometimes unable to accurately predict the quality of this channel.
%$1$.

The EMA model's ability to exploit generalized training 
is quite impressive. 
Adding more channels to the training database (``all'' condition in the table) only led to small degradation with respect to the case when the model was trained on the channel used for test. 
Performing tests on entirely unseen channels (i.e., not present in the training database) made worsening more evident, but the model remained suitable for many application contexts. 
This shows that training EMA on the specific target operating conditions is not strictly mandatory, easing the integration of this model into real devices.

Results for the hourglass and pyramid NN models are reported in Table~\ref{tab:res_NN}. 
They show that NNs were generally able to predict link quality with higher accuracy than the EMA model
(this is particularly true for channels $5$ and $9$). 
The performance of these two models is mostly comparable, but the pyramid NN is seemingly less stable in some specific operating conditions.
See, e.g., the MSE for channel $1$ when training is performed on all the other channels ($\overline{\mathrm{ch1}}$), which is $2.45 \cdot 10^{-3}$ for the hourglass and $3.27 \cdot 10^{-3}$ for the pyramid.
Both NN models maintain good generalization capability, which means that limited performance degradation is expected when non-specific training is carried out (``all'' or, to remove any possible dependency, ``all except the channel under test'').
This is essential when the prediction model has to be included in real equipment, 
where a specific training procedure can hardly be carried out online after deployment.
From this viewpoint, results are generally better than those provided by the EMA.

The above results suggest that feeding the outcomes of multiple EMA filters with different cutoff frequencies to a neural network constitutes a promising approach for link quality prediction,
even when channel-independent training is exploited.
Applying ML techniques to smoothened EMA results (instead of using them directly) provides several advantages, including better accuracy and reliability, due to a deeper understanding of the intrinsic nature of the wireless spectrum seen by \mbox{Wi-Fi} equipment. 
The obvious drawback of NN-based models is their higher computational complexity, for both the training and testing phases. 
Consequently, the choice of the prediction model that best suits any given wireless application, among those proposed in this work, must be made taking into account the trade-off between the ability to provide forecasts as accurately as possible and computational demands.

\section{Conclusions}
\label{sec:conclusions}
In this study, we delved into the crucial domain of wireless link quality prediction, a pivotal aspect for ensuring the seamless operation of communication systems in dense industrial environments, especially within the framework of Industry 4.0. 
The challenges posed by factors like shadowing, multipath propagation, interference, network load, and other environmental complexities necessitate innovative approaches for accurate predictions.

Our investigation encompassed three models, namely EMA, hourglass NN, and pyramid NN,
which were suitably trained before use.
Experimental results revealed notable distinctions in their performance, shedding light on their efficacy in predicting the FDR of a Wi-Fi link in a time interval located in the immediate future.
While exhibiting commendable accuracy in certain scenarios, the EMA model revealed some limitations, particularly in capturing complex patterns inherent in dense indoor environments. 
The hourglass NN and pyramid NN models, leveraging the power of neural networks, showcased superior capabilities in understanding intricate relationships within the data used for training. 
Key insights from our analysis include the superior accuracy demonstrated by the NN models compared to the EMA model. 
Both the NN models featured a heightened understanding of the nuanced factors influencing channel quality, reflecting their potential to improve communication reliability in the dynamic landscapes of Industry 4.0.

As we move forward, the findings from this study advocate for the adoption of neural network-based approaches, such as the hourglass and pyramid models, in wireless link quality prediction tasks. 
Further research could explore optimizations, adaptability to diverse environments, and considerations about real-time implementation, fostering the evolution of communication systems to meet the tight demands of Industry 4.0. 
The ongoing integration of wireless technologies in industrial applications amplifies the significance of accurate channel quality predictions, making advancements in this domain pivotal for the seamless and resilient connectivity demanded by the next industrial revolution.

\bibliographystyle{IEEEtran}
\bibliography{bibliography}

\end{document}